\RequirePackage{ifpdf}
\ifpdf 
\documentclass[pdftex]{sigma}
\else
\documentclass{sigma}
\fi

\begin{document}
\allowdisplaybreaks

\renewcommand{\PaperNumber}{040}

\FirstPageHeading

\ShortArticleName{Nonclassical Approximate Symmetries of Evolution
Equations with a Small Parameter}

\ArticleName{Nonclassical Approximate Symmetries\\ of Evolution
Equations with a Small Parameter}

\Author{Svetlana KORDYUKOVA}

\AuthorNameForHeading{S. Kordyukova}

\Address{Department of Mathematics, Ufa State Aviation Technical University,\\
 12 K.~Marx Str., Ufa, 450000 Russia}
\Email{\href{mailto:sveta.kor05@mail.ru}{sveta.kor05@mail.ru}}

\ArticleDates{Received November 30, 2005, in f\/inal form March
17, 2006; Published online April 10, 2006}

\Abstract{We introduce a method of approximate nonclassical
Lie--B\"acklund symmetries for
 partial dif\/ferential equations with a small parameter and discuss applications
  of this method to f\/inding of approximate solutions both
  integrable and nonintegrable equations.}


\Classification{58J70; 35Q53}

\section{Introduction}

The theory of one- and multi-parameter approximate transformation
groups was initiated by Ibragimov, Baikov, Gazizov \cite{1,12}.
They introduced the notion of approximate Lie--B\"acklund symmetry
of a partial dif\/ferential equation with a small parameter
$\varepsilon$ and develop a method, which allows to construct
approximate Lie--B\"acklund symmetries of such an equation (a
perturbed equation) in the form of a power series in
$\varepsilon$, starting from an exact Lie--B\"acklund symmetry of
the unperturbed equation (for $\varepsilon=0$). Similar ideas were
suggested independently by Fushchych and Shtelen (see, for
instance, \cite{5} and the bibliography therein). The main purpose
of this paper is to extend these methods to approximate
nonclassical Lie--B\"acklund symmetries.

Nonclassical symmetries appeared for the f\/irst time in the paper
by Bluman and Cole in 1969~\cite{2}. Since then this theory was
actively developed in papers of: Olver and Rosenau \cite{3}
(nonclassical method), Clarkson and Kruskal \cite{4} (nonclassical
symmetry reductions (direct method)), Fushchych's school (\cite{5}
and the bibliography therein) (conditional symmetries and
reductions of partial dif\/ferential equations), Fokas and Liu
\cite{6} (the generalized conditional symmetry method), Olver
\cite{10} (nonclassical and conditional symmetries). Nonclassical
Lie--B\"acklund symmetries for evolution equations were considered
in the paper by Zhdanov \cite{7}. This paper also contains
a~theorem on reduction of an evolution equation to a system of
ordinary dif\/ferential equations. The notion of nonclassical
Lie--B\"acklund symmetry is a very wide generalization of the
notion of point symmetry. Nevertheless, in many cases,
nonclassical Lie--B\"acklund symmetries enable to construct
dif\/ferential substitutions, which reduce a partial
dif\/ferential equation to a system of ordinary dif\/ferential
equations. This fact is used for f\/inding new solutions of
partial dif\/ferential equations, which cannot be found with the
help of the classical symmetry method.

The method of approximate conditional symmetries for partial
dif\/ferential equations with a~small parameter was suggested by
Mahomed and Qu~\cite{8} (point symmetries), Kara, Mahomed and Qu
(potential approximate symmetries) \cite{9}. In this paper we
develop the method of approximate nonclassical {\em
Lie--B\"acklund} symmetries. In \cite{1}, Baikov, Gazizov and
Ibragimov constructed approximate Lie--B\"acklund symmetries of
the Korteweg--de Vries equation $u_t=uu_x+\varepsilon u_{xxx}$,
starting from exact symmetries of the transport equation
$u_t=uu_x$. In this paper, we extend this construction to
approximate nonclassical Lie--B\"acklund symmetries.

We will consider a particular class of evolution partial
dif\/ferential equations with a small parameter given by
\[
u_t=uu_x+\varepsilon H(t, x, u, u_x, u_{xx}, \ldots).
\]
This class contains both integrable and nonintegrable equations.
We consider such equations as perturbations of the transport
equation $u_t=uu_x$ and construct approximate nonclassical
symmetries of these equations, starting from exact nonclassical
symmetries of the transport equation. Using these approximate
nonclassical symmetries and the reduction theorem, we f\/ind
approximate conditionally invariant solutions of equations under
consideration. As an example, we f\/ind approximate solutions of
the KdV equation with a small parameter and of some nonintegrable
equations.

\section[Nonclassical Lie-B\"acklund symmetries]{Nonclassical Lie--B\"acklund symmetries}

Recall the def\/inition of classical Lie--B\"acklund symmetries
(here we will consider symmetries given by canonical
Lie--B\"acklund operators):
\begin{definition}
An operator
\begin{gather*}
X=\zeta\frac{\partial}{\partial
u}+(D_x\zeta)\frac{\partial}{\partial
u_x}+(D_t\zeta)\frac{\partial}{\partial
u_t}+(D_{xx}\zeta)\frac{\partial}{\partial u_{xx}}+\cdots,
\end{gather*} where
\[\zeta=\zeta(t,x,u,u_x,u_{xx},\ldots),
\]
will be called a classical Lie--B\"acklund symmetry for a partial
dif\/ferential equation of evolution type
\begin{gather*}
u_t=F(t,x,u,u_x,u_{xx},\ldots),
\end{gather*}
 if
\begin{gather}
\label{opr} X(u_t-F)\big|_{u_t=F}=0.
\end{gather}
\end{definition}

Here $D_x$ and $D_t$ are the total dif\/ferentiation operators:
\begin{gather*}
D_x=\partial_x+\partial_u u_x+\partial_{u_x}
u_{xx}+\partial_{u_{xx}} u_{xxx}+\cdots,
\\
D_t =\partial_t+\partial_u u_t+\partial_{u_x}
u_{xt}+\partial_{u_{xx}} u_{xxt}+\cdots,
\end{gather*}
$D_{xx}=D_x^2=D_x(D_x)$, $D_{xxx}=D_x^3=D_x(D_{xx})$ etc. The
equation (\ref{opr}) is the determining equation for
Lie--B\"acklund symmetries.
\begin{definition}
An operator
\begin{gather}
\label{Xot} X=\eta\frac{\partial}{\partial
u}+(D_x\eta)\frac{\partial}{\partial
u_x}+(D_t\eta)\frac{\partial}{\partial
u_t}+(D_{xx}\eta)\frac{\partial}{\partial u_{xx}}+\cdots,
\end{gather}
where $\eta=\eta(t, x, u, u_x, u_{xx}, \ldots)$, will be called a
nonclassical Lie--B\"acklund symmetry for a~partial dif\/ferential
equation
\[
u_t=F(x,u,u_x,u_{xx},\ldots),
\]
if
\begin{gather}
\label{opr1} X (u_t-F)\Big|_{\mbox{\scriptsize $\begin{array}{@{}l} u_t=F \\
\eta=0\end{array}$}}=0.
\end{gather}
\end{definition}

The equation (\ref{opr1}) is the determining equation for
nonclassical Lie--B\"acklund symmetries. This def\/inition is well
known and can be found in the paper by Zhdanov \cite{7}.

Theory of approximate point symmetries was developed by Baikov,
Gazizov, Ibragimov  in~\cite{1,12}. They proposed to consider
point symmetries in the form of formal power series
\[
X=\overset{0}X+\varepsilon \overset{1}X+\cdots +\varepsilon^n
\overset{n}X+\cdots.
\]
 Now we introduce
approximate nonclassical Lie--B\"acklund symmetries.

\begin{definition}
An operator
\begin{gather}
 X=\left(\sum_{i=0}^n \varepsilon^i
\overset{i}\eta\right)\frac{\partial}{\partial
u}+\left(D_x\left(\sum_{i=0}^n \varepsilon^i
\overset{i}\eta\right)\right)\frac{\partial}{\partial
u_x}+\left(D_t\left(\sum_{i=0}^n \varepsilon^i
\overset{i}\eta\right)\right)\frac{\partial}{\partial
u_t}\nonumber\\
\phantom{X=}{}+\left(D_{xx}\left(\sum_{i=0}^n \varepsilon^i
\overset{i}\eta\right)\right)\frac{\partial}{\partial
u_{xx}}+\cdots,\label{OpR}
\end{gather}
where
$\overset{k}{\eta}=\overset{k}{\eta}(t,x,u,u_x,u_{xx},\ldots)$,
$k=1,2,\ldots,n$ will be called an approximate nonclassical
Lie--B\"acklund symmetry  (in the nth order order of precision)
for an evolution partial dif\/ferential equation with a small
parameter:
\[
u_t=F(t,x,u,u_x,u_{xx},\ldots)+\varepsilon
G(t,x,u,u_x,u_{xx},\ldots)+o(\varepsilon)
\]
if
\begin{gather}
\label{opr2} X (u_t-F-\varepsilon G)\Big|_{\mbox{\scriptsize
$\begin{array}{@{}l} u_t=F+\varepsilon G\\ \sum\limits_{i=0}^n
\varepsilon^i
\overset{i}\eta=o(\varepsilon^n)\end{array}$}}=o(\varepsilon^n).
\end{gather}

The equation (\ref{opr2}) is the determining equation for
approximate nonclassical Lie--B\"acklund symmetries.
\end{definition}

Recall that, by def\/inition, the equality $\alpha(z,\varepsilon)=
o(\varepsilon^p)$ is equivalent to the following condition:
\[
\lim_{\varepsilon\rightarrow
0}\frac{\alpha(z,\varepsilon)}{\varepsilon^p}= 0.
\]
  Here $p$ is
called the order of precision.

We will use the following theorem on stability of symmetries of
the transport equation \cite{1}.

\begin{theorem}[Baikov, Gazizov, Ibragimov]
Any canonical Lie--B\"acklund symmetry
\[
\overset{0}{X}=\overset{0}{\eta}\frac{\partial}{\partial u}+
\left(D_x\overset{0}\eta\right)\frac{\partial}{\partial
u_x}+\left(D_t\overset{0}\eta\right)\frac{\partial}{\partial
u_t}\] of the equation
\begin{gather}\label{qsq}
u_t=h(u)u_x,
\end{gather}
gives rise to an approximate symmetry of the form \eqref{OpR} of
the equation
\begin{gather}\label{qtq}
u_t=h(u)u_x+\varepsilon H(t,x,u,u_x,u_{xx},\ldots)
\end{gather}
with an arbitrary order of precision in $\varepsilon$.
\end{theorem}

In other words, the equation (\ref{qtq}) approximately inherits
all the symmetries of the equation~(\ref{qsq}).

\section{Approximate conditionally invariant solutions}

Now we introduce the def\/inition of approximate conditionally
invariant solutions:

\begin{definition}
An approximate solution of an equation
\[
u_t=F(t,x,u,u_x,u_{xx},\ldots)+\varepsilon
G(t,x,u,u_x,u_{xx},\ldots)+o(\varepsilon)
\]
written in the form of a formal power series
\[
u=\sum_{i=0}^{\infty}\varepsilon^i \overset{i}u
\]
is called conditionally invariant under an approximate
nonclassical symmetry $X$  (in the nth order order of precision),
given by formula (\ref{OpR}), if
\begin{gather*}
\sum_{j=0}^{\infty}
\overset{j}\eta\left(\sum_{i=0}^{\infty}\varepsilon^i
\overset{i}u\right)=o(\varepsilon^n).
\end{gather*}
\end{definition}

As an example, we consider approximate nonclassical symmetries of
the KdV equation
\begin{gather} \label{kdv}
u_t-u u_x-\varepsilon u_{xxx}=0.
\end{gather}

 Take the exact nonclassical Lie--B\"acklund
symmetry of the transport equation:
\[
\overset{0}X=\overset{0}{\eta}\frac{\partial}{\partial u}+\cdots,
\qquad \overset{0}\eta=u_{xx}. \] It is easy to check that this is
not a classical Lie--B\"acklund symmetry.

The corresponding approximate nonclassical Lie--B\"acklund
symmetry of the approximate KdV equation (\ref{kdv}) is written in
the form
\begin{gather}
 X=\Big(\overset{0}\eta+\varepsilon
\overset{1}\eta\Big)\frac{\partial}{\partial u}+
\Big(D_x\Big(\overset{0}\eta+\varepsilon
\overset{1}\eta\Big)\Big)\frac{\partial}{\partial
u_x}+\Big(D_t\Big(\overset{0}\eta+\varepsilon
\overset{1}\eta\Big)\Big)\frac{\partial}{\partial
u_t}\nonumber\\
\phantom{X=}{} +\Big(D_{xxx}\Big(\overset{0}\eta+\varepsilon
\overset{1}\eta\Big)\Big)\frac{\partial}{\partial
u_{xxx}}.\label{Kop}
\end{gather}

From the determining equation (\ref{opr2}) for $X$,  it follows
that
\begin{alignat}{3}
& \varepsilon^0 :\  &&  \overset{0}\eta=u_{xx},&\nonumber\\
& \varepsilon^1 : \ && {\frac {\partial }{\partial
t}}\overset{1}\eta-u_x \overset{1}\eta- u{\frac{\partial
}{\partial x}}\overset{1}\eta+u_{x}^{2}{\frac{
\partial }{\partial u_x}}\overset{1}\eta+3u_x u_{xx}{\frac{\partial }{
\partial u_{xx}}}\overset{1}\eta & \nonumber\\
&&&+\left(3u_{xx}^{2}+4u_xu_{xxx}\right){\frac{
\partial }{\partial u_{xxx}}}\overset{1}\eta +\left(10u_{xx} u_{xxx}+5u_x u_{xxxx}\right ){
\frac{\partial }{\partial u_{xxxx}}}\overset{1}\eta&\nonumber\\
&&& +\left(10u_{xxx}^{2}+15 u_{xx}u_{xxxx}+6u_x u_{xxxxx}\right
){\frac {\partial }{\partial u_{xxxxx}}}\overset{1}\eta
-u_{xxxxx}=0.& \label{et1}
\end{alignat}
Whence we get
 \begin{gather*}  \overset{1}\eta=-F\Bigg(u,x+ut,{\frac {u_xt+1}{u_x}},-{\frac
{u_{xx}}{u_x^{3}}},{\frac {u_x u_{xxx}-3{u_xx}^{2}}{u_x^{5}}},
-{\frac {u_{xxxx}u_x^{2}-10u_x
u_{xx} u_{xxx}+ 15u_{xx}^{3}}{u_x^{7}}},\\
\phantom{\overset{1}\eta=-F\Bigg(}{}-{\frac{105u_{xx}^{4}-u_{xxxxx}u_x^{3}-105u_{xx}^{2}u_x
u_{xxx}+
15u_{xx} u_{xxxx}u_x^{2}+10u_x^{2}u_{xxx}^{2}}{u_x^{9}}}\Bigg)u_x\\
\phantom{\overset{1}\eta=-F\Bigg(}{}+\frac 14{\frac
{u_{xxxxx}}{u_x}}  +{ \frac{-\frac 12 u_{xxx}^{2}-\frac 34
u_{xx}u_{xxxx}}{u_x^{2}}}+\frac 74 {\frac
{u_xx^{2}u_{xxx}}{u_x^{3} }}-\frac 34 {\frac
{u_{xx}^{4}}{u_x^{4}}},
\end{gather*}
where $F$ is an arbitrary function.

 Note that the order of $\overset{1}\eta$ equals the sum of
the orders of $\overset{0}\eta$ and the perturbation $G$ minus
one. Here we consider an approximate conditionally invariant
solution of the KdV equation (\ref{kdv}) in the form:
\[
u=u^0+\varepsilon u^1+o(\varepsilon).
\]
Conditional invariance under an approximate nonclassical symmetry
(\ref{Kop}) in the f\/irst order of precision is written as
\begin{gather}
\label{sol} \overset{0}{\eta}\left(u^0+\varepsilon
u^1\right)+\varepsilon \overset{1}{\eta}\left(u^0+\varepsilon
u^1\right)=o(\varepsilon).
\end{gather}

To compute an approximately invariant solution in the zero order
of precision, we use the following reduction theorem \cite{7}.

\begin{theorem}
Suppose that an equation
\[
u_t=F(t,x,u,u_x,u_{xx},\ldots,u_N),\qquad u_N=\frac{\partial^N
u}{\partial x^N}
\]
is conditionally invariant under a Lie--B\"acklund operator
\eqref{Xot}. Let
\[
u=f(t,x,C_1,C_2,\ldots,C_N)
\]
be a general solution of the equation $\eta(t, x, u, u_1, \ldots,
u_N)=0$. Then the Ansatz
\[
u=f\big(t,x,\varphi_1(t),\varphi_2(t),\ldots,\varphi_N(t)\big),
\]
where $\varphi_j(t)$, $j=1,2,\ldots,N,$ are arbitrary smooth
functions, reduces the partial differential equation $u_t=F$ to a
system of $N$ ordinary differential equations for the functions
$\varphi_j(t)$, $j=1, 2, \ldots, N$.
\end{theorem}

There is a nice consequence of this theorem.

\begin{corollary}
Suppose an equation
\[ u_t=F(t,x,u,u_x,u_{xx},\ldots)+\varepsilon
G(t,x,u,u_x,u_{xx},\ldots)+o(\varepsilon)
\]
admits an approximate Lie--B\"acklund operator $X$, given by the
formula \eqref{OpR} with $n=1$. Let
\[
\overset{0}u=f(t,x,C_1,C_2,\ldots, C_N), \qquad
\overset{1}u=g(t,x,C_1,\ldots,C_{N+M})
\]
be a general solution of the equation
\[
\big(\overset{0}\eta+\varepsilon\overset{1}\eta\big)\big(\overset{0}u+\varepsilon
\overset{1}u\big)=o(\varepsilon).
\]
Then the Ansatz
\begin{gather*}
u=f\big(t,x,\varphi_1(t),\varphi_2(t),\ldots,\varphi_N(t)\big)\\
\phantom{u=}{}+\varepsilon
g\big(t,x,\varphi_1(t),\varphi_2(t),\ldots,\varphi_N(t),
\psi_1(t),\psi_2(t),\ldots,\psi_M(t)\big),
\end{gather*}
reduces the equation $u_t=F+\varepsilon G$ into a system of $N+M$
ordinary differential equations for~$\varphi_j(t)$, $j=1, 2,
\ldots, N,$ and $\psi_k(t)$, $k=1, 2, \ldots, M$.
\end{corollary}

\begin{example} Take a nonclassical symmetry
(\ref{Xot}) ($\eta=\overset{0}\eta$) of the transport equation
\begin{gather} \label{eqtr} u_t=uu_x,
\end{gather}
 where\[ \overset{0}{\eta}=u_{xx}.
\]
Applying the operator (\ref{Xot}) to the equation (\ref{eqtr}) in
the zero order of precision, we have  $\overset{0}{\eta}(u^0)=0,$
whence we get $u^0 = Ax+B.$

By Reduction Theorem, we substitute
\[
u^0 = A(t)x+B(t)
\]
to the transport equation (\ref{eqtr}) and get
\[
\dot{A}= A^2, \qquad \dot{B}=A B.
\]
A general solution has the form:
\[
A=-\frac{1}{t+a}, \qquad B=\frac{b}{t+a},
\]
where $a$, $b$ are constants.

 Thus we get
\[
u^0=\frac{b-x}{t+a}.
\]

Take $\overset{1}{\eta}$ as in (\ref{et1}) with
\[
 F(u)=p\ e^u.
\]
where $p$ is a constant. From (\ref{sol})  it follows that
\[
u^1_{xx} -\frac{p}{t+a} e^{\frac{b-x}{t+a}}=0
\]
and
\[
u^1=p(t+a) e^{\frac{b-x}{t+a}}+C x+D.
\]
Take the approximate solution
\[
u= \frac{b-x}{t+a}+\varepsilon\left(p(t)(t+a)
e^{\frac{b-x}{t+a}}+C(t)x+D(t)\right)
\]
and substitute it into the KdV equation (\ref{kdv}). We get three
f\/irst order ODE for $C(t)$, $D(t)$, $p(t)$:
\[ \dot{C}=-\frac{2
C}{t+a}, \qquad \dot{D}=\frac{b C-D}{t+a}, \qquad
\dot{p}=\frac{-2p}{t+a}.
\]
A general solution of the system can be written as
\[
C(t)=\frac{c_3}{b(t+a)^2},   \qquad D(t)=\frac{c_2 t+c_2 a+c_3
}{(t+a)^2}, \qquad p(t)=\frac{c_1}{(t+a)^2}
\]
where $c_1$, $c_2$, $c_3$ are constants.

Finally, we get the following solution of the KdV equation in the
f\/irst order of precision:
\[
u= \frac{b-x}{t+a}+\varepsilon\left(\frac{c_1}{t+a}
e^{\frac{b-x}{t+a}}+\frac{c_3 x}{b(t+a)^2}+\frac{c_2 t+c_2 a+c_3
}{(t+a)^2}\right).
\]
\end{example}

We use the following proposition to construct nonclassical
symmetries.

\begin{proposition}\label{p:1}
Let
\[
X=\eta\frac{\partial}{\partial u}+\cdots, \qquad
\eta=\eta(t,x,u,u_x,u_{xx},\ldots),
\]
be a classical Lie--B\"acklund symmetry for a f\/irst order PDE
\begin{gather}\label{eb:too}
F(t,x,u,u_x,u_t)=0.
\end{gather}
For any function $f=f(t,x,u,u_x,u_{xx},\ldots)$, the operator
\[
\overset{*}{X}=\overset{*}{\eta}\frac{\partial}{\partial
u}+\cdots, \qquad \overset{*}{\eta}=f\eta,
\]
is a nonclassical Lie--B\"acklund symmetry for \eqref{eb:too}.
\end{proposition}

\begin{example}\label{example}
Now we consider an example of f\/inding symmetries of the KdV
equation with a~small parameter (\ref{kdv})  and construct its
approximate solution. We have a classical Lie--B\"acklund symmetry
\[
X=\overset{0}{\eta}\frac{\partial}{\partial
u}+D_x\big(\overset{0}{\eta}\big)\frac{\partial}{\partial
u_x}+D_t\big(\overset{0}{\eta}\big)\frac{\partial}{\partial u_t}
\]
 of the
transport equation (\ref{eqtr}), where
\[
\overset{0}{\eta}=u_x \Phi\left(u,
x+ut,\frac{u_xt+1}{u_x},-\frac{u_{xx}}{u_x^3},
\frac{u_xu_{xxx}-3u_{xx}^2}{u_x^5}\right).
\]
 By Proposition~\ref{p:1},
$\overset{0}{\eta}=u_xu_{xxx}-3u_{xx}^2$ is a nonclassical
Lie--B\"acklund symmetry of the transport equation (\ref{eqtr}).
Now we take  operator (\ref{Kop}).
 Applying the operator $X$  to the equation (\ref{kdv}),  we get the following
 equations in the zero and f\/irst
 orders of precision in $\varepsilon$:
\begin{alignat*}{3}
&\varepsilon^0 : && \ \overset{0}\eta=u_xu_{xxx}-3u_{xx}^2,&\\
& \varepsilon^1 : \ & & {\frac {\partial }{\partial
t}}\overset{1}\eta-u_x \overset{1}\eta-u{\frac {\partial
}{\partial x}}\overset{1}\eta+u_x^{2}{ \frac {\partial }{\partial
u_x}}\overset{1}\eta+3 u_x u_{xx}{\frac {
\partial }{\partial u_{xx}}}\overset{1}\eta  +(3 u_{xx}^{2}+4 u_x
u_{xxx}){\frac {\partial }{\partial u_{xxx}}}\overset{1}\eta&\\
&&& {}+(10  u_{xx} u_{xxx} +5 u_x u_{xxxx}){\frac {\partial
}{\partial u_{xxxx}}}\overset{1}\eta  + (10 u_{xxx}^{2}+15 u_{xx}
u_{xxxx}+6 u_x u_{xxxxx}){\frac {\partial }{\partial u_{xxxxx}}}
\overset{1}\eta& \\
&&& {} +(35 u_{xxx} u_{xxxx}+21 u_{xx} u_{xxxxx}  +7 u_x
u_{xxx}){\frac {
\partial }{\partial u_{xxxxxx}}}\overset{1}\eta+14 u_{xxx} u_{xxxx}&\\
&&& {}+3 u_{xx} u_{xxxxx}-u_x u_{xxxxxx}=0.&
\end{alignat*}

From the last equation, we f\/ind
\begin{gather*}
\overset{1}\eta =-F\Bigg(u,x+ut,{\frac {u_x t+1}{u_x}},-{\frac
{u_{xx}}{ u_x^3}},{\frac {u_x u_{xxx}-3 u_{xx}^2}{u_x^5}},-{\frac
{15 u_{xx}^3+u_{xxxx}
u_x ^2-10 u_{xx} u_x u_{xxx} }{u_x^7}},\\
\phantom{\overset{1}\eta =}{} -{\frac {105
u_{xx}^4-u_{xxxxx}u_x^3-105 u_{xx}^2u_x u_{xxx}+ 15 u_{xx}
u_{xxxx} u_x^2+10 u_x^2u_{xxx}^2}{u_{xxx}^9}},
\\
\phantom{\overset{1}\eta =}{} -{\frac{945 u_{xx}^5-1260
u_{xx}^3u_x u_{xxx} +280 u_{xx}u_x^2u_{xxx}^2+210
u_{xx}^2u_{xxxx}u_x^2
-21 u_{xx} u_{xxxxx}u_x^3}{u_x^{11}}}\\
\phantom{\overset{1}\eta =}{} +\frac{-35  u_x^3 u_{xxx} u_{xxxx}
+u_{xxxxxx}u_x^4}{u_x^{11}}\!\Bigg)u_x +\frac 16
u_{xxxxxx}+\!\left(\!-{\frac {13}{14}} u_{xx } u_{xxxxx}\!-
\!{\frac {17}{6}} u_{xxx} u_{xxxx}\! \right)\!u_x^{-1}\\
\phantom{\overset{1}\eta =}{} + \left({\frac {395}{84}}
u_{xx}u_{xxx}^2+{ \frac {157}{56}}
u_{xx}^2u_{xxxx}\right)u_x^{-2}-{\frac {25}{4}} {\frac
{u_{xx}^3u_{xxx}}{u_x^3}}+{\frac {15}{8}} {\frac
{u_{xx}^5}{u_x^4}},
\end{gather*}
where $F$ is an arbitrary function. The invariance condition of a
solution
\[
u=\overset{0}u+\varepsilon \overset{1}u+\cdots
\]
 in the
f\/irst order of precision is written as
\begin{gather} \label{equ1}
\big(\overset{0}\eta+\varepsilon
\overset{1}\eta\big)\big(\overset{0}u+\varepsilon
\overset{1}u\big)=o(\varepsilon).
\end{gather}
If we substitute $\overset{0}\eta$, $\overset{1}\eta$ to the
equation (\ref{equ1}), we obtain in the zero and f\/irst orders of
precision by~$\varepsilon$ equations for $\overset{0}u$ and
$\overset{1}u$:
\begin{alignat*}{3}
&\varepsilon^0  : \ && \overset{0}\eta\big(\overset{0}u\big)=0,&\\
&\varepsilon^1  : \ && \overset{1}u_x
\overset{0}u_{xxx}+\overset{0}u_x \overset{1}u_{xxx}-6
\overset{1}u_{xx}
\overset{0}u_{xx}+\overset{1}\eta\big(\overset{0}u\big)=0.&
\end{alignat*}

 We f\/ind
\[
\overset{0}u=2 \sqrt{t^2+t-x}-2t-1
\]
 and substitute it in the
second equation:
\begin{gather}
\label{eq11} -\frac{\overset{1}u_{xxx}}{\sqrt{t^2+t-x}}-\frac{3
\overset{1}u_x}{4(t^2+t-x)^{5/2}}+3\frac{
\overset{1}u_{xx}}{(t^2+t-x)^{3/2}}+c(t^2+t-x)^{-11/2}=0,
\end{gather}
where $c$ is a constant, depending on the choice of  $F$. The
equation (\ref{eq11}) is an ordinary dif\/ferential equation and
has the following solution:
\begin{gather*}
\overset{1}u=\frac{2 c
}{15(t^2+t-x)^{2}}+F_1(t)+\frac{F_2(t)}{\sqrt{t^2+t-x}}
+F_3(t)\sqrt{t^2+t-x}.
\end{gather*}
If we substitute  $u=\overset{0}u+\varepsilon \overset{1}u$ in
(\ref{kdv}) we obtain a system of ordinary dif\/ferential
equations for f\/inding $F_1(t)$, $F_2(t)$, $F_3(t)$:
\[\dot{F_1}=-2F_3, \qquad  \dot{F_2}=-F_1,  \qquad  \dot{F_3}=0,
\]
which has the solution:
\[
F_1=-2At+B, \qquad   F_2=At^2-Bt+C, \qquad F_3=A,
\]
where $A$, $B$, $C$ are  arbitrary constants, $c=\frac{1}{4}$.
Finally, we f\/ind the solution of (\ref{kdv}):
\begin{gather*}
u=2\sqrt{t^2+t-x}-2t-1\\
\phantom{u=}{}+\varepsilon\left(\frac{1}{4}(t^2+t-x)^{
2}+(-2At+B)+ \frac{At^2-Bt+C}{\sqrt{t^2+t-x}}
+A\sqrt{t^2+t-x}\right),
\end{gather*}
where $A$, $B$, $C$ are arbitrary constants.
\end{example}

\begin{example}
Now we consider an example of f\/inding of symmetries of the
nonintegrable equation
\begin{gather}
\label{eq34} u_t=uu_x+u_{xxx}^2
\end{gather}
and construct its approximate solution. Using the criteria of
integrability, it can be checked that the equation (\ref{eq34}) is
nonintegrable  \cite{11}.

As in Example~\ref{example}, take a nonclassical Lie--B\"acklund
symmetry of the transport equation (\ref{eqtr}) with
$\overset{0}{\eta}=u_xu_{xxx}-3u_{xx}^2$.
 Applying  the operator, given by (\ref{Kop}), to the equation (\ref{eq34})
 we get in the zero and f\/irst
 orders of precision by $\varepsilon$:
\begin{alignat*}{3}
& \varepsilon^0 : \ & & \ \overset{0}\eta=u_xu_{xxx}-3u_{xx}^2,\\
& \varepsilon^1 : & & {\frac {\partial }{\partial
t}}\overset{1}\eta-u_x \overset{1}\eta-u{\frac {\partial
}{\partial x}}\overset{1}\eta+u_x^{2}{ \frac {\partial }{\partial
u_x}}\overset{1}\eta+3 u_x u_{xx}{\frac {
\partial }{\partial u_{xx}}}\overset{1}\eta +\left (3 u_{xx}^{2}+4 u_x
u_{xxx} \right ){\frac {\partial }{\partial
u_{xxx}}}\overset{1}\eta&\\
&&& {} + (10  u_{xx} u_{xxx}+5 u_x u_{xxxx} )\frac {\partial
}{\partial u_{xxxx}}\overset{1}\eta + (10 u_{xxx}^{2}+ 15 u_{xx}
u_{xxxx}+6 u_x u_{xxxxx}){\frac {\partial }{\partial u_{xxxxx}}}
\overset{1}\eta&\\
&&& {}+(35 u_{xxx} u_{xxxx}+21 u_{xx} u_{xxxxx} +7 u_x
u_{xxx}){\frac {
\partial }{\partial u_{xxxxxx}}}\overset{1}\eta&\\
&&& +2 u_{xxx}(14 u_{xxx} u_{xxxx} +3 u_{xx} u_{xxxxx}-u_x
u_{xxxxxx})=0.&
\end{alignat*}

From the last equation, we f\/ind
\begin{gather*}
\overset{1}\eta =-F\Bigg(u,x+ut,{\frac {u_x t+1}{u_x}},-{\frac
{u_{xx}}{ u_x^{3}}},{\frac {u_x u_{xxx}-3 u_{xx}^{2}}{u_x^{5}}},
-{\frac {15 u_{xx}^{3}+u_{xxxx} u_x ^{2}-10 u_{xx} u_x u_{xxx}
}{u_x^{7}}}, \\
\phantom{\overset{1}\eta =}{} -{\frac {105
u_{xx}^{4}-u_{xxxxx}u_x^{3} -105 u_{xx}^{2}u_x u_{xxx}+ 15 u_{xx}
u_{xxxx}
u_x^{2}+10 u_x^{2}u_{xxx}^{2}}{u_{xxx}^{9}}},\\
\phantom{\overset{1}\eta =}{} -{\frac {945 u_{xx}^{5}-1260
u_{xx}^{3}u_x u_{xxx} +280 u_{xx}u_x^{2}u_{xxx}^{2}+210
u_{xx}^{2}u_{xxxx}u_x^{2}
-21 u_{xx} u_{xxxxx}u_x^{3}}{u_x^{11}}}\\
\phantom{\overset{1}\eta =}{}+\frac{{-35
 u_x^{3}u_{xxx} u_{xxxx} +u_{xxxxxx}u_x^{4}}}{u_x^{11}}\Bigg)u_x
+\frac{1}{5}u_{xxx} u_{xxxxxx}\\
\phantom{\overset{1}\eta =}{}-\left(\frac{51}{55} u_{xxx}u_{xx}
u_{xxxxx}-\frac{3}{55}u_{xx}^2 u_{xxxxxx}-\frac{35}{11}u_{xxx}^2
u_{xxxx}\right) u_x^{-1}\\
\phantom{\overset{1}\eta =}{} +\left(\frac{32}{11}u_{xx}^2 u_{xxx}
u_{xxxx}
+\frac{113}{33}u_{xx}u_{xxx}^3+\frac{18}{55}u_{xx}^3u_{xxxxx}\right)
u_x^{-2}\\
\phantom{\overset{1}\eta =}{} +\left(-\frac{695}{143}u_{xx}^3
u_{xxx}^2-\frac{150}{143}u_{xx}^4 u_{xxxx}\right)u_x^{-3}
+\frac{405}{143}u_{xx}^5 u_{xxx} u_x^{-4}-\frac{81}{143} u_{xx}^7
u_x^{-5}.
\end{gather*}
where $F$ is an arbitrary function. Now we f\/ind an approximate
solution of the equation (\ref{kdv}) in the form
$u=\overset{0}u+\varepsilon \overset{1}u+o(\varepsilon)$. The
invariance condition in the f\/irst order of precision is written
as:
\begin{gather} \label{equ11}
\big(\overset{0}\eta+\varepsilon
\overset{1}\eta\big)\big(\overset{0}u+\varepsilon
\overset{1}u\big)=o(\varepsilon).
\end{gather}
If we substitute $\overset{0}\eta$, $\overset{1}\eta$ to the
equation (\ref{equ11}), we obtain in the zero and f\/irst orders
by $\varepsilon$  equations for $\overset{0}u$ and $\overset{1}u$:
\begin{alignat*}{3}
& \varepsilon^0 : \ && \overset{0}\eta\big(\overset{0}u\big)=0,&\\
& \varepsilon^1 : \ && \overset{1}u_x
\overset{0}u_{xxx}+\overset{0}u_x \overset{1}u_{xxx}-6
\overset{1}u_{xx}
\overset{0}u_{xx}+\overset{1}\eta\big(\overset{0}u\big)=0.
\end{alignat*}

From the f\/irst equation, we get
\[
\overset{0}u=2 \sqrt{t^2+t-x}-2t-1,
\]
and, substituting this expression into the second equation, we
obtain
\begin{gather}
\label{trt}
 -\frac{\overset{1}u_{xxx}}{\sqrt{t^2+t-x}}-\frac{3
\overset{1}u_x}{4(t^2+t-x)^{5/2}}+3\frac{
\overset{1}u_{xx}}{(t^2+t-x)^{3/2}}+c\left(t^2+t-x\right)^{-8}=0,
\end{gather}
where $c$ is a constant depending on the choice of $F$. The
equation (\ref{trt}) is an ordinary dif\/ferential equation and
has the following solution:
\begin{gather}
\label{eq23}
\overset{1}u=\frac{c}{90}\left(t^2+t-x\right)^{-9/2}+(-2At+B)+
\frac{At^2-Bt+C}{\sqrt{t^2+t-x}} +A\sqrt{t^2+t-x}.
\end{gather}
If we substitute  $u=\overset{0}u+\varepsilon \overset{1}u$ in the
equation (\ref{eq34}) we obtain the system of ordinary
dif\/ferential equations for f\/inding $F_1(t)$, $F_2(t)$,
$F_3(t):$
\[\dot{F_1}=-2F_3, \qquad  \dot{F_2}=-F_1,  \qquad  \dot{F_3}=0,
\]
which has the solution:
\[
F_1=-2At+B, \qquad   F_2=At^2-Bt+C, \qquad F_3=A.
\]
Therefore, the solution $u$ has the form:
\begin{gather*}
u=2\sqrt{t^2+t-x}-2t-1\\
\phantom{u=}{}
+\varepsilon\left(\frac{405}{64}\left(t^2+t-x\right)^{-9/
2}+(-2At+B)+ \frac{At^2-Bt+C}{\sqrt{t^2+t-x}}
+A\sqrt{t^2+t-x}\right),
\end{gather*}
where $A$, $B$, $C$ are arbitrary constants.
\end{example}

\begin{remark}
One can show that the approximate symmetries constructed in the
above examples remain stable in any higher order of precision.
However, we do not know whether any non-classical symmetry of an
evolution partial dif\/ferential equation with a small parameter
is stable in any order of precision.
\end{remark}

\section{Conclusion}
The methods developed in this paper can be applied to larger
classes of partial dif\/ferential equations with a small
parameter, not only to the evolution ones. For instance, in the
paper~\cite{BK03} it is shown that classical approximate
Lie--B\"acklund symmetries of the Boussinesq equation with a small
parameter can be constructed, starting from the exact
Lie--B\"acklund symmetries of the linear wave equation. It is
quite possible that these results can be extended to non-classical
approximate symmetries of the Boussinesq equation.

From the other side, one should note that stability property of
approximate classical symmetries holds only for a very restricted
class of partial dif\/ferential equations with a small parameter,
mainly, for those, which have very nice symmetry properties in the
zero order of precision. The class of non-classical symmetries is
much larger than the class of classical symmetries. Therefore, one
can hardly expect to have some general theorems on stability of
non-classical symmetries. This means that we will have to
investigate separately stability properties of non-classical
symmetries in each particular case.

All the computations have been made with the help of Maple.

\LastPageEnding
\end{document}